%% file: neurips_2026.tex
\documentclass{article}

 \usepackage[preprint]{neurips_2026}
\usepackage{neurips_2026}
\usepackage{graphicx}
\usepackage{subcaption}
\usepackage{amsmath}
\usepackage{mathalpha}
\usepackage{pifont}
\usepackage{multirow}
\usepackage{colortbl}
\usepackage{makecell}
\usepackage{algorithm}
\usepackage{algorithmic}
\usepackage{tabularx}
\usepackage{enumitem}

\usepackage[utf8]{inputenc} % allow utf-8 input
\usepackage[T1]{fontenc}    % use 8-bit T1 fonts
\usepackage{hyperref}       % hyperlinks
\usepackage{url}            % simple URL typesetting
\usepackage{booktabs}       % professional-quality tables
\usepackage{amsfonts}       % blackboard math symbols
\usepackage{nicefrac}       % compact symbols for 1/2, etc.
\usepackage{microtype}      % microtypography
\usepackage{xcolor}         % colors

\setcitestyle{numbers}
\setcitestyle{numbers,square}

\title{Seeing Inside the Storm: Improving Nowcasting by Integrating Meteorological Drivers}

\author{%
  Minghui Qiu\\
  HKUST(GZ)\\
  \And
  Jun Chen\\
  HKUST(GZ)\\
  \AND
  Lin Chen \\
  HKUST(GZ)\\
  \And
  Weifeng Chen \\
  HKUST(GZ)\\
  \And
  Shuxin Zhong \\
  HKUST(GZ)\\
   \And
   Zhidan Liu \\
  HKUST(GZ)\\
   \And
  Yu Zhang \\
 Guangzhou Meteorological Observatory\\
   \And
  Kaishun Wu \\
  HKUST(GZ)\\
}

\begin{document}

\maketitle

\input{sec/8cmd}

% \begin{document}
% \maketitle
\input{sec/0abstract}
\input{sec/1intro}
\input{sec/2relatedwork}

\input{sec/3preliminaries}
\input{sec/4framework}

\input{sec/5experiments}

\input{sec/6discussion}
\input{sec/7conclusion}

% %%%%%%%%%%%%%%%%%%%%%%%%%%%%%%%%%%%%%%%%%%%%%%%%%%%%%%%%%%%%
\bibliographystyle{plainnat}
\bibliography{main}

\appendix

\section{Experiment Setting Details}

\subsection{Detailed Dataset Configurations}
\label{sec:appendix_dataset}
In our experiments, we utilize the \textbf{3D-NEXRDA} dataset~\cite{Datasets}, a large-scale benchmark designed for high-resolution 3D radar observations of severe convective storms across the continental United States. The temporal coverage of the data spans from January 2020 to December 2022. 

To comprehensively capture the complex physical dynamics of storm events, the dataset includes 7 core radar variables. Each storm event is structured as a 25-frame radar sequence covering a duration of 125 minutes. Spatially, the data features a horizontal grid resolution of approximately $2 \text{ km} \times 2 \text{ km}$, and a vertical resolution comprising 28 levels that range from 0.5 km to 22 km in altitude.

Specifically, these variables capture distinct yet interconnected atmospheric properties, providing a full-spectrum view of convective systems:
\begin{itemize}[leftmargin=*]
    \item \textbf{Horizontal Reflectivity ($Z_H$):} The primary indicator of precipitation intensity and storm morphology, essential for tracking the core structure and development of convective systems.
    \item \textbf{Differential Reflectivity ($Z_{DR}$):} Measures the reflectivity difference between horizontal and vertical polarizations, providing critical insights into hydrometeor shape and size (e.g., distinguishing large raindrops from hail).
    \item \textbf{Specific Differential Phase ($K_{DP}$):} Captures the propagation phase difference, which is highly sensitive to liquid water content and invaluable for identifying areas of intense rainfall.
    \item \textbf{Correlation Coefficient ($\rho_{HV}$):} Assesses the consistency of hydrometeor shapes and orientations within a radar volume, helping to identify melting layers and distinguish meteorological from non-meteorological echoes (e.g., tornado debris).
    \item \textbf{Velocity Spectrum Width ($SW$):} Represents the variance of radial velocities within a sampling volume, serving as a proxy for sub-grid scale turbulence and wind shear within the storm.
    \item \textbf{Azimuthal Shear ($AzShr$):} Measures the gradient of radial velocity in the azimuthal direction, a key kinematic feature for detecting rotating updrafts (mesocyclones) and tornadic signatures.
    \item \textbf{Radial Divergence ($Div$):} Captures the radial gradient of velocity, which helps characterize the strength of storm top divergence (updrafts) and low-level outflows (cold pools).
\end{itemize}

\subsection{Data Splits and Leakage Control.} 
\label{split}
The dataset division is strictly based on chronological periods: the training and verification sets are drawn from 2020 to 2021, whereas the test set spans 2022. To strictly prevent temporal leakage, samples are strictly grouped and isolated by contiguous temporal sequences.

\subsection{Baseline Details}
\label{baseline}
\subsubsection{Baselines}
We compare \N\ with nine competitive models.
\begin{itemize}[leftmargin=*]
    \item \textbf{ConvLSTM}~\cite{shi2015convolutional} models radar evolution as a sequence-to-sequence video prediction task to capture local spatiotemporal dependencies. 
    \item \textbf{PhyDNet}~\cite{guen2020disentangling} introduces physics-inspired inductive bias through separate motion and content encoders, enabling better physical interpretability.
    \item \textbf{PredRNN-V2}~\cite{wang2022predrnn} builds upon ConvLSTM by introducing spatiotemporal memory cells, improving long-range dependency modeling.
    
    \item \textbf{Earthformer}~\cite{gao2022earthformer} uses cuboid attention to capture short-range spatiotemporal dependencies.  
    \item \textbf{Earthfarseer}~\cite{wu2024earthfarsser} extends Earthformer with causal far-sight modules and temporal fusion. 
    \item \textbf{WF-UNet}~\cite{kaparakis2023wf} employs modality-specific encoding with two separate CNN-based encoders for wind and radar inputs, followed by late fusion near the decision layer.
    \item \textbf{SimVP}~\cite{gao2022simvp} decouples spatial encoding and temporal modeling into separate stages through lightweight convolutional architectures.  
   \item \textbf{DGMR}~\cite{ravuri2021skilful} employs a conditional generative adversarial network (GAN) to produce probabilistic forecasts, overcoming the blurring effect of traditional deterministic approaches to generate realistic, spatio-temporally consistent radar sequences.  
\item \textbf{NowcastNet}~\cite{zhang2023skilful} unifies a physical evolution model for advection with a deep generative network, specifically designed to capture multi-scale dynamics and significantly improve the prediction of extreme precipitation events.
    \item \textbf{PastNet}~\cite{wu2024pastnet} leverages a multi-path encoder with physics-informed inductive biases, aiming to balance forecasting accuracy and computational efficiency. 
    \item \textbf{AlphaPre}~\cite{Lin_2025_CVPR} improves spatial detail preservation by disentangling amplitude and phase components in the frequency domain.
\end{itemize} 

\subsection{Evaluation Metrics: Definitions and Formulations}
\label{appendix:metrics}

To evaluate both the accuracy and physical realism of~\N's predictions, we employ a comprehensive set of metrics:

\begin{itemize} [leftmargin=*]
    \item \textbf{Mean Absolute Error (MAE)} measures per-pixel numerical deviation, where \textit{lower values indicate more accurate forecasting}. \\
    $$ \text{MAE} = \frac{1}{N} \sum_{i=1}^{N} \left| y_i - \hat{y}_i \right| $$
    where $y_i$ is the ground truth, $\hat{y}_i$ is the predicted value, and $N$ is the total number of pixels.

    \item \textbf{Critical Success Index (CSI)} computed at 20, 30, and 40 dBZ thresholds to assess~\N's ability to detect storms of varying severity, where \textit{higher CSI values indicate improved detection of high-impact convective cores with fewer false alarms.} \\

    $$ \text{CSI} = \frac{\text{TP}}{\text{TP} + \text{FN} + \text{FP}} $$
    where TP (True Positives), FN (False Negatives), and FP (False Positives) are calculated pixel-wise based on the specified dBZ thresholds.

    \item \textbf{Structural Similarity Index (SSIM)} evaluates the preservation of storm morphology and structural fidelity, where \textit{higher values imply more coherent and visually plausible storm structures.} \\
    
    $$ \text{SSIM}(x, y) = \frac{(2\mu_x\mu_y + C_1)(2\sigma_{xy} + C_2)}{(\mu_x^2 + \mu_y^2 + C_1)(\sigma_x^2 + \sigma_y^2 + C_2)} $$
    where $\mu$ and $\sigma$ denote the mean and variance/covariance of the true ($x$) and predicted ($y$) images, and $C_1, C_2$ are small constants to maintain numerical stability.

    \item \textbf{Signal-to-Noise Ratio (SNR)} evaluates textural sharpness and the ratio of the meaningful signal to prediction error, where \textit{higher values imply more coherent and visually plausible storm structures.} \\
    
    $$ \text{SNR} = 10 \log_{10} \left( \frac{\sum_{i=1}^{N} y_i^2}{\sum_{i=1}^{N} (y_i - \hat{y}_i)^2} \right) $$
    where $y_i$ represents the original ground truth signal, and $y_i - \hat{y}_i$ represents the noise introduced by the prediction error.
\end{itemize}

\subsubsection{Implementation Details}
\label{Implementation}
\N\ is trained on a single NVIDIA A800 GPU and a Intel(R) Xeon(R) Platinum 8358P CPU @ 2.60GHz for 200 epochs with a batch size of 8. 
We set the number of warmup epochs to 20. The weight $\lambda_s$ of the pretext loss is set to 1 during the warmup phase and is fixed at $1 \times 10^{-4}$ thereafter.
We adopt the Adam optimizer with the learning rate set to $1 \times 10^{-3}$.

\section{Extra Experiment Results}

\subsection{Statistical Significance Analysis}
\label{sec:statistical}
To ensure empirical rigor and reproducibility, all evaluations are conducted across five independent random seeds. We assess the performance gains of~\N\ against the best-performing baseline using the Wilcoxon signed-rank test, a non-parametric method robust to non-normal data distributions. The results demonstrate that~\N's improvements are consistent across all runs and statistically significant ($p < 0.05$). This confirms that our reported advancements stem from the architectural strengths of~\N\ rather than stochastic variation, ensuring that the model's superiority is both robust and statistically sound.

\begin{figure}[h]
  \centering
  \includegraphics[width=0.8\linewidth]{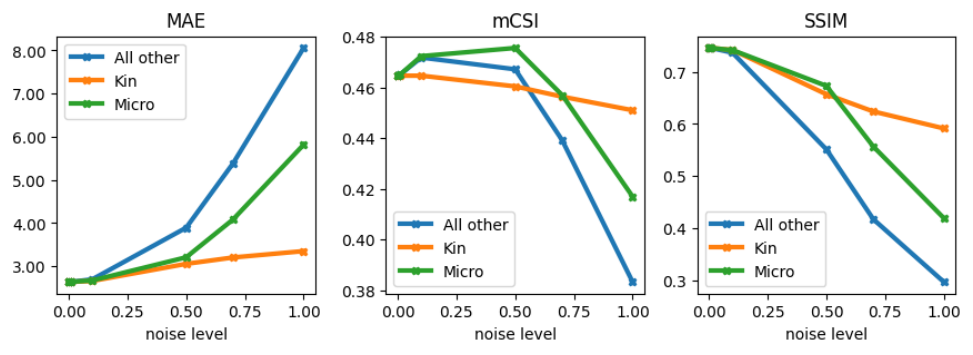}
  \caption{Results of the degradation study, showing model performance under varying levels of additive Gaussian noise.}
  \label{fig:deg}
\end{figure}

\subsection{Robustness and Degradation Study}
\label{sec:robustness}
To evaluate the robustness of our model against noisy inputs, we conducted a degradation study by applying varying levels of additive Gaussian noise to the auxiliary meteorological inputs during inference. As illustrated in Figure~\ref{fig:deg}, the model's performance decays gradually rather than catastrophically as noise levels increase. This indicates that \N~ effectively leverages the auxiliary meteorological drivers for enhancement without collapsing or over-relying on them.

\subsection{Additional Ablation Studies}
\label{sec:additional_ablation}
We performed supplementary ablation studies to further validate the effectiveness of the proposed modules: (a) replacing the \textit{Temporal-Phase Aligner} with a standard concatenation (Add-up) operation, and (b) replacing the \textit{Cross-Field Spatial Aggregator} with standard attention (w/o Condition). 

As shown in Table~\ref{tab:ablation-components}, removing the condition in the \textit{Aggregator} significantly drops performance, highlighting the importance of successfully incorporating slow-varying background information. Meanwhile, the \textit{Temporal-Phase Aligner} primarily boosts the SSIM score, validating its value in modeling temporal lead-lag stages and structural fidelity.

\begin{table}[h]
\centering
\caption{Additional ablation studies.}
\resizebox{0.8\linewidth}{!}{
\setlength{\tabcolsep}{1.2mm}
\begin{tabular}{lccccccc}
\toprule
\textbf{Variant} & \textbf{MAE↓} & \textbf{mCSI↑} & \textbf{CSI$_{20}$↑} & \textbf{CSI$_{30}$↑} & \textbf{CSI$_{40}$↑} & \textbf{SNR↑} & \textbf{SSIM↑} \\
\midrule
Add-up & 2.664 & \textbf{0.433} & \textbf{0.608} & \textbf{0.490} & \textbf{0.200} & 6.142 & 0.738 \\
w/o Cond. & 2.698 & 0.424 & 0.605 & 0.482 & 0.184 & 6.133 & 0.733 \\
\midrule
\textbf{Ours} & \textbf{2.660} & 0.431 & 0.607 & 0.488 & 0.198 & \textbf{6.183} & \textbf{0.738} \\
\bottomrule
\end{tabular}
}

\label{tab:ablation-components}
\end{table}

\subsection{Efficiency and Resolution Scaling}
\label{sec:efficiency}

\paragraph{Computational Efficiency Benchmarks.} 
We benchmarked the inference frames per second (FPS), parameter counts, and floating-point operations (FLOPs) against NowcastNet and Earthformer under a $128\times128$ resolution. Table~\ref{tab:Benchmarks} demonstrates that \N~ maintains a balanced trade-off, offering competitive inference speeds and manageable computational overhead while achieving superior forecasting accuracy. Note that for fairness, baselines evaluated under MISO/MIMO were retrained by adjusting the input channels while maintaining a comparable computational load.

\begin{table}[h]
\centering
\caption{Efficiency comparison of different models at $128\times128$ resolution.}
\resizebox{0.7\linewidth}{!}{
\begin{tabular}{lccc}
\toprule
\textbf{Model} &\textbf{Inference FPS} & \textbf{Params (M)} & \textbf{FLOPs (G)} \\
\midrule
\textbf{Ours (MIMO)} &14.63  & 13.03 & 88.47 \\
\textbf{Ours (MISO)} &13.69 & 12.93 & 87.97 \\
DGMR (MISO) &1.41 & 15.40 & 19.56 \\
NowcastNet (MISO) &40.36 & 35.10 & 18.75 \\
Earthformer (MISO) &102.56 & 5.05 & 104.00 \\
\bottomrule
\end{tabular}
}

\label{tab:Benchmarks}
\end{table}

\paragraph{Resolution Scaling.} 
To demonstrate the scalability of \N~, we report inference speed (Latency) and memory consumption across varying spatial resolutions ($64\times64$, $128\times128$, and $256\times256$). As detailed in Table~\ref{tab:resolution_inference}, \N~ maintains acceptable inference latency and favorable memory scaling characteristics even at higher resolutions.

\begin{table}[h]
\centering
\caption{Inference speed and memory consumption under different resolutions.}
\resizebox{0.8\linewidth}{!}{
\begin{tabular}{lccc}
\toprule
\textbf{Resolution} & \textbf{Latency (ms)} & \textbf{Peak Mem. (MB)} & \textbf{Dynamic Mem. (MB)} \\
\midrule
\multicolumn{4}{c}{\textbf{Multi-Input Single-Output (MISO)}} \\
\midrule
$64\times64$   & 68.67  & 235.85  & 188.90 \\
$128\times128$ & 73.02  & 1139.27 & 1089.26 \\
$256\times256$ & 152.64 & 8237.44 & 8175.24 \\
\midrule
\multicolumn{4}{c}{\textbf{Multi-Input Multi-Output (MIMO)}} \\
\midrule
$64\times64$   & 68.32  & 236.23  & 188.90 \\
$128\times128$ & 76.50  & 1139.65 & 1089.26 \\
$256\times256$ & 154.79 & 8237.82 & 8175.24 \\
\bottomrule
\end{tabular}
}

\label{tab:resolution_inference}
\end{table}

\subsection {Visualization Results}
\label{vis}
This section includes visualization results across different baselines and forecasting horizons, which have been discussed in the main text. Conventional methods often produce overly smoothed predictions and exhibit spatial inconsistencies at longer lead times, whereas \N\ better preserves the structural evolution and intensity of convective systems. These additional examples further demonstrate the robustness and physical consistency of \N\ in modeling storm dynamics.
\begin{figure}[h!]
\centering
% \caption{Sensitivity Analysis}
\includegraphics[width=1\linewidth]{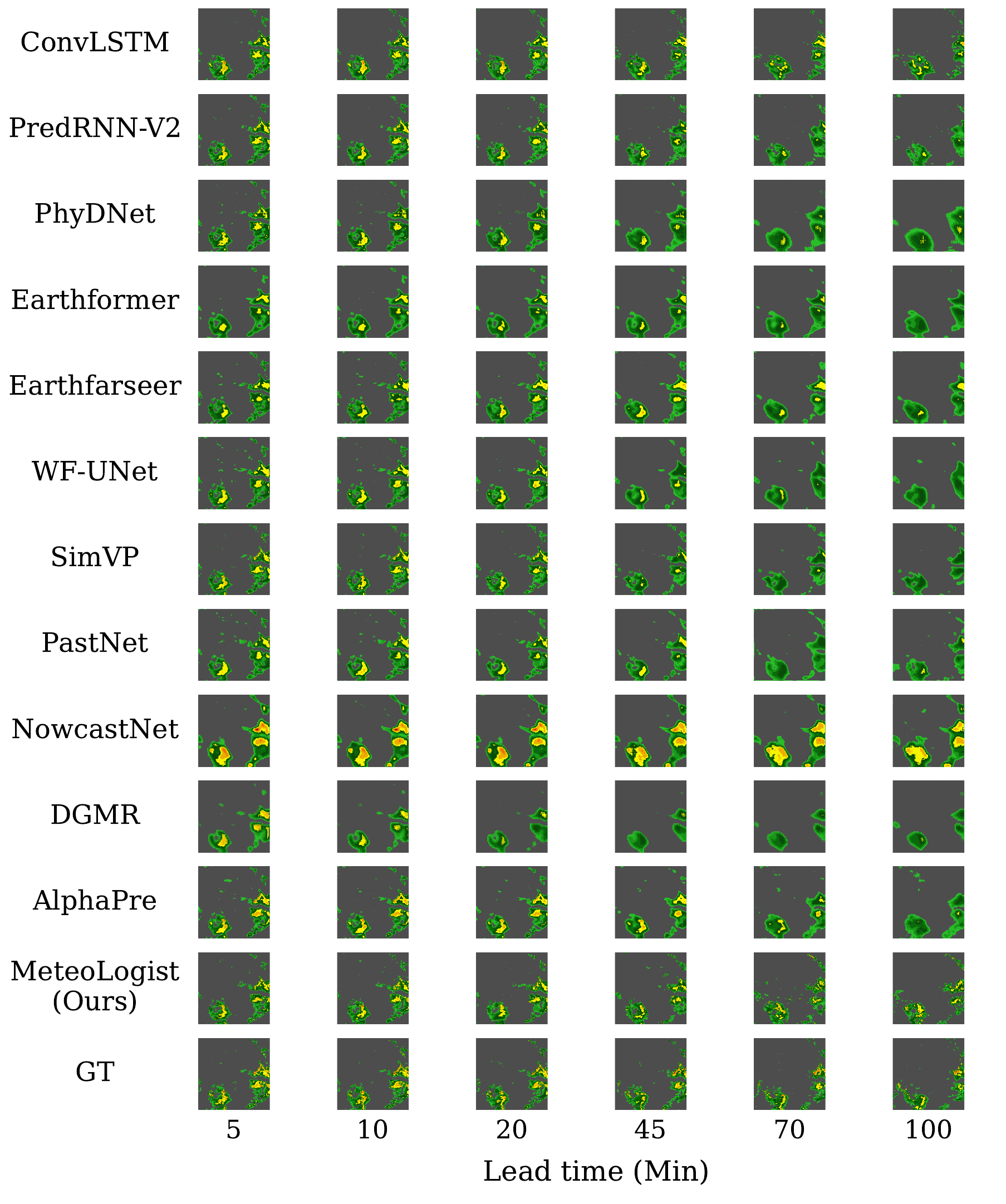}
\caption{
Qualitative Comparison of Radar Reflectivity Forecasts across Models at Increasing Lead Times.  
}
\label{fig:case}
\vspace{-1em}
\end{figure}

\section{Discussion}

\label{sec:discussion}
% In this section, we summarize some insights and lessons learned. 
% We also discuss the limitations and future directions, followed by implication and generalization. 

\subsection{Insights and Lessons Learned}
Our comprehensive analysis yields two key insights:
\begin{itemize}[leftmargin=*]
    \item \textbf{Multi-input fusion is not always beneficial—it depends on architecture.}
    Simply adding radar variables may hurt performance, especially in recurrent models (supported by Table~\ref{tab:SISO&MISO-performance}).
    Effective fusion requires inductive bias to preserve cross-channel semantics—a gap \N\ explicitly addresses.
    \item \textbf{Physically disentangled design yields generalization and interpretability.}  
    By explicitly modeling microphysical and kinematic components, \N\ achieves better performance (supported by Table~\ref{tab:ablation}), offering a more interpretable and extensible forecasting backbone. 

    % \item \textbf{Bidirectional correction between radar and surface stations improves both spatial precision and semantic fidelity.}  
    % Radar provides dense spatial structure but suffers from semantic artifacts; surface stations offer sparse but reliable truth. 
    % Treating them as mutually corrective—rather than independent—modalities enables the model to localize accurately and anchor meaningfully, leading to more coherent reconstructions. 
    % % (\textit{supported by Figure~\ref{}}).

    % \item \textbf{Rainfall regimes demand adaptive spatiotemporal reasoning.}  
    % Sensitivity studies reveal that optimal receptive fields and memory depths vary with rainfall intensity: light rain favors broad spatial and temporal context, whereas heavy rain hinges on localized, short-term cues. 
    % This underscores the need for regime-aware architecture that adjusts to physical heterogeneity.
\end{itemize}

\subsection{Limitations and Future Work}
While~\N\ performs well across tasks, it relies on radar inputs and hand-crafted physical encoders. 
Future work will explore end-to-end learning of physical abstractions and adaptation to broader modalities (e.g., satellite, surface stations).
Moreover, the current framework focuses on a single vertical layer, limiting its ability to model vertically coupled convective processes.
Future work will extend~\N\ to volumetric prediction to better align with the 3D propagation characteristics of severe storms.

\subsection{Implication and Generalization}
\N's physically disentangled design naturally generalizes to critical meteorological tasks—such as storm lifecycle classification, convective event detection, and precipitation phase estimation—that are directly tied to real-world societal impacts, such as earlier hazard warnings.
\begin{itemize}[leftmargin=*]
    
\item \textbf{\N\ is inherently extensible toward probabilistic methods.} While the current version operates as a deterministic backbone emphasizing physically grounded representations, this solid foundation enables seamless integration with generative paradigms. Specifically, the physically informed encoders of \N\ can serve as interpretable priors for diffusion-based generative models, thereby facilitating calibrated uncertainty estimation and advancing reliable probabilistic precipitation nowcasting.
\item \textbf{\N\ is designed to balance accuracy and operational feasibility.} On a single NVIDIA A800 GPU, the model processes a 128×128 radar frame in 132 ms for single-step forecasting and 134 ms for 20-frame sequence prediction. The peak memory usage is 1.1 GB, with a static footprint of only 50 MB, which is modest compared to other radar nowcasting models and fully compatible with standard operational hardware. 
% \N\ therefore meets sub-minute nowcasting requirements and remains lightweight relative to its accuracy gains.
\end{itemize}

\subsection{Social Impact}
This work contributes to the growing intersection of machine learning and climate adaptation. By shifting the focus of precipitation nowcasting from merely extrapolating visible radar echoes to proactively anticipating convective initiation, \N~ provides critical lead time for extreme weather early warning systems. In the face of increasing global climate volatility, extending this intervention window is vital for disaster preparedness, aviation safety, and urban flood management.

Despite these positive societal implications, deploying deep learning models in safety-critical meteorological domains introduces inherent risks. Primarily, data-driven models are susceptible to out-of-distribution shifts.
As climate change rapidly alters historical weather patterns, \N~ may encounter unprecedented atmospheric dynamics, potentially leading to false alarms (false positives) or missed detections of severe storms (false negatives). False positives can cause unnecessary economic and societal disruption, eventually eroding public trust in automated warning systems, while false negatives carry severe consequences for human safety. Additionally, deploying such systems risks automation bias, where human forecasters might over-rely on the AI's predictions, neglecting to scrutinize the underlying meteorological context.

To mitigate these risks, \N~ intentionally integrates physical inductive biases—explicitly disentangling thermodynamic, kinematic, and microphysical streams—which significantly enhances model transparency and interpretability compared to standard black-box video predictors. For responsible operational deployment, we strongly advocate utilizing \N~ strictly as a decision-support tool within a human-in-the-loop framework. It should complement,

\end{document}

%% file: sec/8cmd.tex
\newcommand{\N}{\texttt{MeteoLogist}}

\newcommand{\note}{\color{red}}

\newcommand{\ComponentA}{Physics-Tailored Encoders} % Physical-Disentangled Perception
\newcommand{\ComponentB}{Temporal-Phase Aligner}
\newcommand{\ComponentC}{Cross-Field Spatial Aggregator}

\newcommand{\AsubComponentA}{Thermodynamic Encoder}
\newcommand{\AsubComponentB}{Kinematic Encoder}
\newcommand{\AsubComponentC}{Microphysical Encoder}

% \newcommand{\AsubComponentB}{Kinematic Encoder}
% \newcommand{\AsubComponentC}{Microphysical Encoder}

% \newcommand{\BsubComponentA}{Temporal Logic Assembler}
% \newcommand{\BsubComponentB}{Spatial Cascade Extractor}

%% file: sec/0abstract.tex
\begin{abstract}
% The June 2025 Guizhou flood, which claimed six lives within minutes, exposed this vulnerability.

% When storms erupt, every second counts—yet most nowcasting systems, built on radar observations, reveal only the current precipitation pattern, not the atmospheric processes that precede it.
% Before any echoes appear, however, the atmosphere already whispers its intent through low-level convergence, turbulent eddies, and latent heating.
% Capturing these ephemeral cues opens a narrow but vital window to move beyond motion tracking toward storm-birth prediction.
Most nowcasting systems, built on radar reflectivity, focus on current precipitation, ignoring the atmospheric precursors—such as low-level convergence, turbulent eddies, and latent heating—that offer a fleeting window to foresee storm birth.
We introduce~\N, a physics-inspired radar intelligence framework that models the full life cycle of convection—from its precursors to organized storm evolution.
However, exploiting these precursors is non-trivial:
they originate from multiple meteorological drivers—thermodynamic, kinematic, and microphysical—that evolve asynchronously (\textbf{C1}) and remain spatially fragmented (\textbf{C2}).
To this end, \N\ designs three tightly integrated components.
The \textbf{\ComponentA} process radar echoes according to their intrinsic physical scales and semantics, forming thermodynamic, kinematic, and microphysical streams that capture distinct dynamical regimes.
The \textbf{\ComponentB} addresses~\textbf{C1} by leveraging causal temporal attention to capture when and how different drivers interact and activate. 
The \textbf{\ComponentC} addresses~\textbf{C2} through cross-regional fusion, aligning weak and scattered precursors across neighboring cells to expose upstream triggers and enforce spatial coherence.
Evaluated on 3D-NEXRAD (2020–2022, US-wide), \N\ boosts high-impact detection (CSI$_{40}$) by +9.7\% over strong baselines,
and achieves a remarkable 37.67\% gain during the storm-developing stage—demonstrating true foresight in sensing storms before they appear.
\textbf{The code can be found in the supplementary material.}

% (\textbf{C1}) \textbf{temporal desynchronization of meteorological drivers}—thermodynamic buildup, boundary-layer convergence, and microphysical bursts evolve on distinct time scales, rarely peaking together;
% and
% (\textbf{C2}) \textbf{spatial disorganization of convective energy}—precursors are weak and scattered across cells, and only through their emergent, non-local alignment does deep convection ignite. 

% The June 2025 Guizhou flood exposed a fatal weakness in modern nowcasting: 
% forecasters failed to catch the moment convection ignited, and the window for action vanished.

% But storms do not erupt from nothing; 
% they build silently through a cascade of physical drivers: thermodynamic instability, dynamic uplift, and rapid microphysical shifts.
% This reality demands a shift—from tracking echoes to anticipating their formation. 

\end{abstract}

%% file: sec/1intro.tex
\section{Introduction}
\label{sec:intro}

Extreme weather leaves little room for delay~\cite{racah2017extremeweather}.
The June 2025 Guizhou flood offered a stark reminder: 
within minutes, 
a localized storm intensified, overwhelming responders and claiming six lives~\cite{Guizhou_floods, Europeanfloods}.
As climate volatility grows and intervention windows shrink~\cite{reichstein2025early}, 
the ability to accurately anticipate convective escalation has become not only a scientific challenge—but a societal imperative~\cite{yano2018scientific}.

At the heart of current nowcasting systems lies radar echo extrapolation, 
which future reflectivity is inferred from past observations using optical flow~\cite{woo2017operational}, ConvLSTM~\cite{shi2015convolutional}, Transformer~\cite{wang2024nuwadynamics}, or diffusion models~\cite{yu2024diffcast}.
Despite effective at tracking echoes displacement, 
these methods are limited by design:
relying solely on reflectivity patterns, they remain blind to convection that has yet to appear~\cite{zhang2023skilful, wu2024earthfarsser}.

Meteorology reveals a deeper story behind what radar sees.
Convection rarely erupts spontaneously—it develops through the coordinated evolution of \textit{thermodynamic instability, kinematic uplift, and microphysical transitions}~\cite{doswell1996flash, schumacher2020formation}. 
Even before the radar echo appears, subtle precursors can be sensed:
low-level convergence in radial velocity, 
turbulent eddies in Doppler spectrum width, 
latent heating in polarimetric variables~\cite{pan2021improving}.
These early signals appear minutes before reflectivity—offering a fleeting yet vital window to \textit{foresee storm birth rather than merely track its motion}.

However, exploiting these precursors introduces two fundamental challenges rooted in atmospheric physics.
\begin{itemize} [leftmargin=*]
    \item \textbf{(C1) 
    The thermodynamic, kinematic, and microphysical processes that drive convection evolve on mismatched time scales}~\cite{fan2018substantial,lee2017interactions}.
    Thermodynamic instability may accumulate slowly for hours before dynamic lift occurs~\cite{walsh2018eight},
    while sharp boundary-layer convergence can ignite convection within minutes under weak instability~\cite{riaz2006onset},
    and microphysical bursts may lead or lag vertical motion by several cycles~\cite{drew2013coalescence}.
    
    \item \textbf{(C2) Convection organizes only when thermodynamic, kinematic, and microphysical processes couple across space}~\cite{zscheischler2020typology}. 
    For instance, localized boundary-layer convergence can trigger moist ascent that releases latent heat aloft, intensifying nearby instability and reorganizing surrounding cells into a coherent convective cluster~\cite{bennett2006review, bauer2015quiet}.

\end{itemize}

To address these challenges, we introduce~\N, a physics-inspired radar intelligence framework that unifies three specialized components to model the full life cycle of convection—from its precursors to organized storm evolution.
First, the \textit{\ComponentA} decomposes radar echoes along intrinsic physical dimensions, enabling process-aware representations:
a~\textit{\AsubComponentA} encodes slowly evolving thermodynamic fields that precondition the environment;
a~\textit{\AsubComponentB} extracts rapid kinematic signatures (e.g., low-level inflows, vertical shear); 
and a~\textit{\AsubComponentC} traces microphysical transitions indicative of early storm formation.
Second,
to address \textbf{C1}, the \textit{\ComponentB} employs a \textit{causal temporal attention mechanism} that models asynchronous activation across drivers—learning the lead-lag dynamics that reveal when multi-physics coupling signals imminent initiation.
Finally, to address \textbf{C2}, the \textit{\ComponentC} introduces a \textit{geo-context attention mechanism} that aggregates non-local cues—restoring mesoscale spatial coherence and tracing upstream triggers that govern storm organization. 
Our main contributions are listed as follows:
\begin{itemize} [leftmargin=*]
    \item 
    While prior nowcasting models extrapolate radar reflectivity, \N\ integrates atmospheric physics—fusing thermodynamic, kinematic, and microphysical cues—to move beyond echo tracking toward storm-birth forecasting.
    
    \item \N\ designs three physics-inspired components:
    \textit{\ComponentA} disentangle radar inputs into process-aware thermodynamic, kinematic, and microphysical streams, 
    \textit{\ComponentB} captures asynchronous, cross-driver interactions to infer when multi-physics coupling triggers convection,
    and \textit{\ComponentC} restores mesoscale spatial coherence via geo-context fusion of upstream precursors.

    \item Evaluated on US-wide 3D-NEXRAD data (2020-2022), \N\ achieves
    +9.705\% improvement in high-impact event detection (CSI$_{40}$),
    +1.761\% in structural fidelity (SSIM),
    and a -2.096\% reduction in intensity error (MAE), along with a remarkable 37.67\% gain during storm development—marking a decisive step toward \textit{sensing storms before they appear}.
    
\end{itemize}

%% file: sec/2relatedwork.tex
\section{Related Work}
\label{sec:related_work}

% We organize related work into two categories: vision-centric echo extrapolation and physics-informed methods. 

\subsection{Video-Centric Echo Extrapolation}
% Radar echo extrapolation is conventionally framed as a video prediction problem-forecasting reflectivity maps frame by frame. 
% Early efforts adopt \textit{recurrent-based models}, modeling temporal dependencies sequentially via hidden states.
% ConvLSTM~\cite{shi2015convolutional} integrates convolutional operations with LSTM cells to capture local spatiotemporal patterns,
% while PredRNN-V2~\cite{wang2022predrnn} introduces spatiotemporal memory to better preserve long-range dependencies.
% Yet, these models are inherently limited in parallelism and struggle to retain sharp spatial details, often producing blurry predictions in high-resolution convective scenes.
% To overcome these bottlenecks, \textit{recurrent-free models} have gained traction.
% SimVP~\cite{gao2022simvp} decouples spatial and temporal modeling through a pure CNN framework,
% and transformer-based models (e.g., Earthformer~\cite{gao2022earthformer} and Earthfarseer~\cite{wu2024earthfarsser}) leverage self-attention to capture global context.
% AlphaPre~\cite{lin2025alphapre} further enhances spatial fidelity by disentangling amplitude and phase in the frequency domain.
% Generative methods such as Diffcast\cite{yu2024diffcast} embed echo evolution into a latent diffusion space,
% generating diverse but plausible future frames via iterative sampling.
% Despite architectural progress, these models remain grounded in a \textit{vision-centric paradigm—treating radar echoes as textures that deform over time}.
% They track where storms move, but miss the deeper question:~\textit{when and why do they form?} 
Radar echo extrapolation is conventionally framed as a video prediction problem. Early \textit{recurrent-based models} (e.g., ConvLSTM~\cite{shi2015convolutional}, PredRNN-V2~\cite{wang2022predrnn}) capture spatiotemporal dependencies but suffer from limited parallelism and blurry outputs in high-resolution convective scenes. To overcome these bottlenecks, \textit{recurrent-free models} have emerged: SimVP~\cite{gao2022simvp} decouples spatiotemporal modeling via CNNs, transformers (Earthformer~\cite{gao2022earthformer}, Earthfarseer~\cite{wu2024earthfarsser}) capture global context, and AlphaPre~\cite{lin2025alphapre} enhances spatial fidelity via frequency disentanglement. Additionally, generative methods like Diffcast~\cite{yu2024diffcast} leverage latent diffusion to sample diverse, plausible futures.
Despite such architectural leaps, these models remain grounded in a \textit{vision-centric paradigm—treating radar echoes merely as textures that deform over time}. They successfully track where storms move, but ignore the underlying physical drivers, missing the deeper question: \textit{when and why do they form?}

\subsection{Physics-Informed Echo Extrapolation}
To move beyond surface-level echo tracking, recent work incorporates physics-informed priors. 
PhyDNet~\cite{guen2020disentangling} separates physical motion from latent content using dual-branch encoders.
NowcastNet~\cite{zhang2023skilful} models echo evolution as an advective process, regularized by a learned residual aligned with the continuity equation.
PastNet~\cite{wu2024pastnet} further embeds physical biases into a multi-path encoder to capture cross-scale signals.
However, they remain focused on refining visible echoes, lacking explicit modeling of the causal cascade by which thermodynamic buildup, dynamic triggers, and microphysical shifts \textit{ignite convection}.

%% file: sec/4framework.tex
\section{Methodology}
\label{sec:method}
%还有很多东西要改，具体看实现，而且需要填充一些内容
\begin{figure*}[t]
  \centering
  \includegraphics[width=1\textwidth]{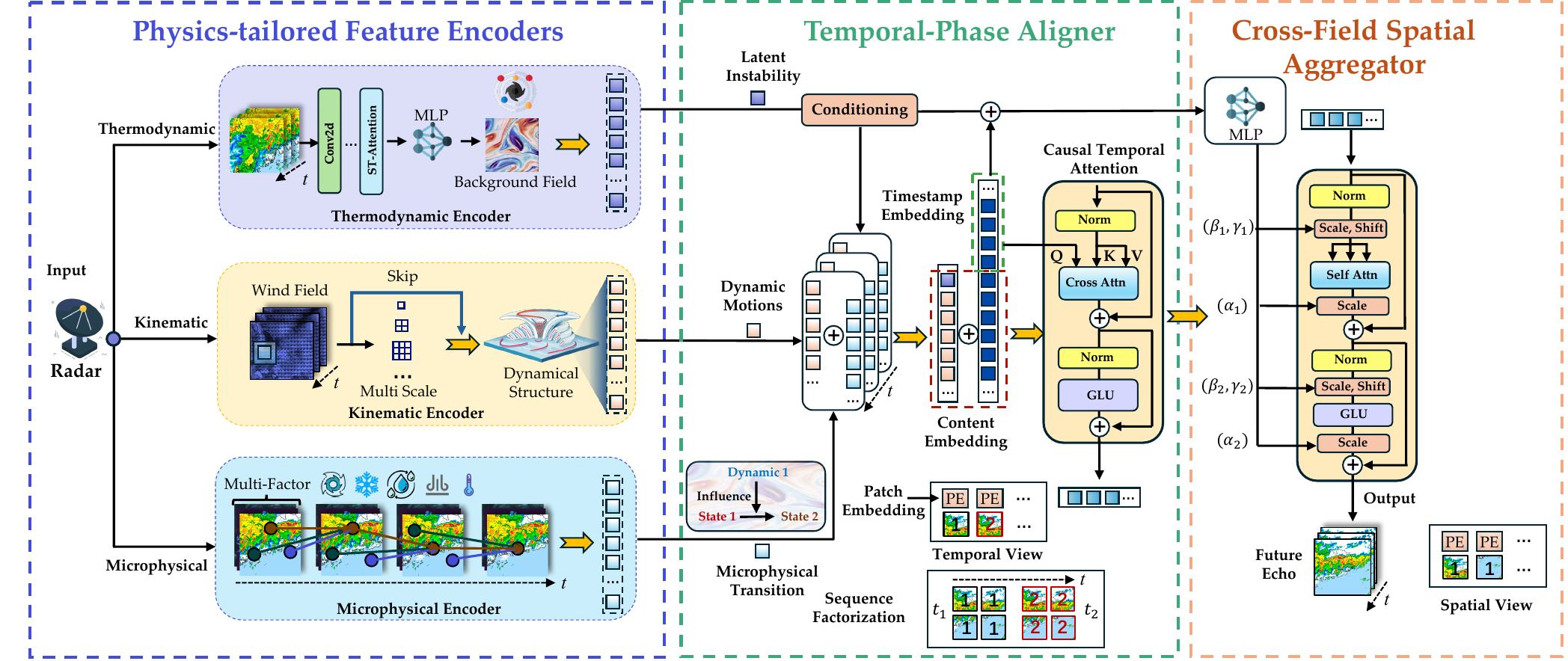}
  \caption{The Framework of~\N. }
  % \vspace{-1em}
  \label{fig:framework}
\end{figure*}

% We first define the radar nowcasting problem, then present the architecture of \N, and finally detail each core component aligned with meteorological processes.

\subsection{Problem Definition}
\label{sec:preliminary}

% We re-conceptualize radar echo extrapolation as a physics-driven reasoning task:
% \textit{not merely predicting where echoes move, but when and why convection initiates}.  
We re-conceptualize radar echo extrapolation as a physics-inspired forecasting task that begins before the echo appears, aiming to capture the precursors that precede convective initiation.
At each time step $t$, a radar observation is represented as a tensor $\mathcal{X}_t \in \mathbb{R}^{H \times W \times C_{\text{in}}}$, 
where $H$ and $W$ are spatial dimensions, 
and $C_{\text{in}}$ is the number of input channels.  
Grounded in meteorological principles, 
we decompose $\mathcal{X}_t$ into three physically meaningful pathways:
(i) \textbf{thermodynamic background ($\mathcal{B}$)}: 
slow-evolving mesoscale conditions (e.g., CAPE, stratification), encoded via $f_{\mathcal{B}}([\mathcal{X}_t]_{t=1}^T)$;
(ii) \textbf{kinematic precursors ($[\mathcal{X}_t^{\text{kin}}]_{t=1}^T$)}: fast-evolving triggers (e.g., low-level convergence or shear), captured via azimuthal shear $AzShr$, radial divergence $Div$, and velocity spectrum width $SW$;
and (iii) \textbf{microphysical signals ($[\mathcal{X}_t^{\text{mic}}]_{t=1}^T$)}: localized particle‐scale transitions (e.g., hydrometeor growth, phase shift) in reflectivity and polarimetric variables (horizontal reflectivity $Z_H$, specific differential phase $K_{DP}$, differential reflectivity $Z_{DR}$, correlation coefficient $\rho_{HV}$).
Our goal is to forecast the next $K$ frames by jointly modeling these three processes: 
\begin{equation}
    [\mathcal{Y}_{T+t}]_{t=1}^K = \mathcal{F}_{\theta}([\mathcal{X}^{\text{mic}}_t]_{t=1}^T, [\mathcal{X}^{\text{kin}}_t]_{t=1}^T; \mathcal{B}),  
\end{equation}
where each output $\mathcal{Y}_{T+t} = \{y_{T+t}^{\mathcal{B}}, y_{T+t}^{\text{kin}}, y_{T+t}^{\text{mic}}\}$ are multi-channel reflectivity field that aligns with thermodynamic, kinematic, and microphysical components. 

% These channels encode the spatial distribution of various meteorological variables, including horizontal reflectivity ($Z_H$), velocity spectrum width ($SW$), azimuthal shear ($AzShr$), radial divergence ($Div$), differential reflectivity ($Z_{DR}$), specific differential phase ($K_{DP}$), and correlation coefficient ($\rho_{HV}$).

\subsection{Overall Architecture}
\N\ is a physics-inspired radar intelligence framework that forecasts not only the motion but also the emergence and evolution of convective storms. 
It consists of three meteorologically grounded components (Figure~\ref{fig:framework}):
\begin{itemize}[leftmargin=*]
    \item \textit{\ComponentA} decompose radar inputs into physically meaningful streams. 
    A \textit{\AsubComponentA} extracts slow-evolving mesoscale processes (e.g., stratification), 
    a \textit{\AsubComponentB} identifies fast-evolving dynamic triggers (e.g., inflow, shear), 
    and a \textit{\AsubComponentC} captures fine-scale hydrometeor transitions (e.g., hydrometeor growth).

    \item \textit{\ComponentB} models asynchronous, cross-driver activation to infer when multi-physics interactions (i.e., thermodynamic buildup, dynamic lift, and microphysical growth) jointly triggers convection. 

    \item \textit{\ComponentC} restores mesoscale spatial coherence by fusing upstream cues across thermodynamic, kinematic, and microphysical fields through a geo-context attention mechanism.
\end{itemize}

\subsection{\ComponentA}

\textit{\ComponentA} decompose radar inputs along the natural chronology of convective development—mirroring \textit{how storms emerge from energy buildup, ignite through dynamic forcing, and materialize as organized precipitation.}
The component comprises three specialized branches:
a~\textit{\AsubComponentA} for slow mesoscale conditioning, 
a~\textit{\AsubComponentB} for rapid uplift triggers, 
and a~\textit{\AsubComponentC} for fine-grained precipitation signals.

\subsubsection{\AsubComponentA}
To model the thermodynamic instability, e.g., mesoscale warming or moistening, that are slow and distributed evolution patterns, we introduce a block-wise self-attention mechanism~\cite{gao2022earthformer}.
Given a radar sequence $[\mathcal{X}_t]_{t=1}^T$, the model first decomposes it into non-overlapping 3D blocks:
\begin{equation}
    \{\mathcal{X}_n\}_{n=1}^N = \texttt{Decompose}([\mathcal{X}_t]_{t=1}^T; b_T, b_H, b_W),
\end{equation}
where $b_T, b_H, b_W$ defines the temporal and spatial granularity.
Within each block, we apply a localized self-attention mechanism~\cite{vaswani2017attention} to extract coherent thermodynamic signals:
\begin{equation}
    \mathcal{X}^{\text{out}}_n = \texttt{BlockAttn}(\mathcal{X}_n) = \text{Softmax}(\frac{Q_n K_n^\top}{\sqrt{d}}) V_n, 
\end{equation}
where $Q_n$, $K_n$, and $V_n$ are learned projections of $\mathcal{X}_n$.
Outputs are reassembled to form a background representation:
\begin{equation}
    \mathcal{B} = \texttt{Merge}(\{\mathcal{X}^{\text{out}}_n\}_{n=1}^N).
\end{equation}
 
\subsubsection{\AsubComponentB}
The motion-sensitive input $\mathcal{X}^{\textbf{kin}}$ includes three channels of data: spectrum width $X^{SW}$, azimuthal shear $X^{AzShr}$ and divergence $X^{Div}$. 
They are measured using Doppler-based method, indicating fast-evolving dynamic triggers (e.g., shear zones, divergent outflows, or gust fronts).
To detect these triggers, we deploy a Multi-Scale Fully Convolutional mechanism~\cite{wu2024earthfarsser} that captures multi-scale kinematic disruptions that signal convective onset.
To be specific, at each time step, we apply depthwise-separable convolutions across multiple receptive fields to capture both fine-grained perturbations and mesoscale convergence:
\begin{equation}
    \mathcal{K}_t = \sum_{k \in \{3,5,7\}} \text{Conv2D}_{k \times k}(\text{Conv2D}_{1 \times 1}(\mathcal{X}^{\textbf{kin}}_t)). 
\end{equation} 

\subsubsection{\AsubComponentC}
The microphysics-aware input $\mathcal{X}^{\textbf{mic}}$ includes data from four different channels: horizontal reflectivity $X^{Z_H}$, differential reflectivity $X^{Z_{DR}}$, specific differential phase $X^{K_{DP}}$, and correlation coefficient $X^{\rho_{HV}}$. 
They are measured using Doppler-polarimetric techniques, representing instant status instead of evolution.
To properly capture the micorphysical process (e.g., hydrometeor growth, melting, and phase transitions), we introduce a flow-guided recurrent encoder design grounded in trajectory-aware warping~\cite{shi2017deep} that could dynamically capture the evolving flow drifts across frames of input. 
Specifically, we use a multi-layer perception network to estimate the motion fields $(\mathcal{U}_t, \mathcal{V}_t) = \texttt{MLP}(X)$.
The motion fields are used to warp the latent feature with bilinear sampling mechanism, enabling the learning of the evolution trajectory.
\begin{equation}
    \mathcal{H}^{(k)}_{t-1} = \sum_{l=1}^L\mathcal{W}^{(l, \,k)}\texttt{warp}\left(\mathcal{M}_{t-1}, \mathcal{U}_t^{(l)}, \mathcal{V}_t^{(l)}\right)
\end{equation}
where the $L$ indicates the number of different trajectories in use. 
The derived hidden states $\mathcal{H}^{(\cdot)}_{t-1}$ are used to compute the microphysical responses embedding $\mathcal{M}_t$ at timestep $t$ with Gated Recurrent Unit (GRU)~\cite{dey2017gate}.
\begin{equation}
\begin{aligned}
    \mathcal{Z}_t &= \sigma \left(\mathcal{W}_{z}  \mathcal{X}^{\text{mic}}_t + \mathcal{H}_{t-1}^{(z)}\right), \\ 
    \mathcal{R}_t &= \sigma \left(\mathcal{W}_{r} \mathcal{X}^{\text{mic}}_t + \mathcal{H}_{t-1}^{(r)}\right), \\ 
    \mathcal{M}'_t &= f \left(\mathcal{W}_{h}\mathcal{X}^{\text{mic}}_t + \mathcal{R}_t \circ \mathcal{H}^{(h)}_{t-1}\right), \\
    \mathcal{M}_t &= (1-\mathcal{Z}_t) \circ \mathcal{M}'_t + \mathcal{Z}_t \circ \mathcal{M}_{i-1}.
\end{aligned}
\end{equation}

\subsection{\ComponentB}
\textit{\ComponentB} learns how physical triggers unfold over time—inferring when and under what conditions disparate signals align to initiate convection.
It consists of two key stages: \textit{capturing local cause-effect transitions} and \textit{projecting them under evolving thermodynamic contexts}.

\subsubsection{Instantaneous Trigger Encoding}
We construct a temporally offset token embedding $\mathcal{E}_{t+1}$, where kinematic signals ($\mathcal{K}_t$) are paired with microphysical responses ($\mathcal{M}_{t+1}$) at the next time step, to capture short-term initiator-response patterns:
\begin{equation}
    % \mathcal{E}_t = \texttt{Concat}(K_t, M_{t+1}).
    \mathcal{E}_{t+1} = \mathcal{K}_{t} + \mathcal{M}_{t+1}
    \label{eq:triggerEnc}
\end{equation}

\subsubsection{Conditional Trigger Projection}
To account for slowly evolving environmental constraints, we condition each trigger $\mathcal{E}_t$ on the thermodynamic background $\mathcal{B}$. 
Precisely, we replace the first input token $\mathcal{E}_1$  with the token computed using $\mathcal{B}$, i.e., $\mathcal{E}'_t \in \{\mathcal{B}, \mathcal{E}_2,\cdots,\mathcal{E}_T\}$. 
The fused context is then encoded via an MLP to yield attention-ready key–value pairs~\cite{chen2021crossvit}:
\begin{equation}
    (K_t^{\text{cond}}, V_t^{\text{cond}}) = \texttt{MLP}(\mathcal{Q}_{t} + \mathcal{E}'_t)
\end{equation}
Future queries attend past triggers to anticipate evolution:
\begin{equation}
    \tilde{\mathcal{Q}}_{T+t'} = \texttt{CrossAttn}(\mathcal{Q}_{T+t'}; \{K_t^{\text{cond}}, V_t^{\text{cond}}\}_{t=1}^{T}).
\end{equation}
Finally, a Gated Linear Unit (GLU) modulates the output to enhance feature selection~\cite{ye2023video}:
\begin{equation}
\mathcal{E}^{CA}_{T+t'} = \texttt{GLU}(\tilde{\mathcal{Q}}_{T+t'}).
\end{equation}

\subsection{\ComponentC}
To enhance spatial coherence and uncover upstream precursors (e.g., cold pools, mesoscale banding),~\textit{\ComponentC} integrates future trigger embeddings with their thermodynamic context and apply geo-context reasoning. 
Specifically, we propose \texttt{Back-AdaLN}, an adaptive layer normalization mechanism with time-evolving thermodynamic background context. 
We fuse the background context token $\mathcal{B}$ with the queried temporal embedding $\mathcal{Q}_{T+t'}$.
\begin{equation}
    \gamma, \alpha, \beta = \texttt{MLP}(\mathcal{B}, \mathcal{Q}_{T+t'})  
\end{equation}
The adaptive layer normalization is defined as,
\begin{equation}
    \texttt{Back-AdaLN} (h , \mathcal{F})= \alpha (\mathcal{F}(\gamma\,\texttt{LayerNorm}(h) + \beta )) + h
\end{equation}
For each queried timestep $T+t'$, ~\textit{\ComponentC} allows queried patterns rendering along with temporal-aware thermodynamic condition,
\begin{equation}
\begin{aligned}
    \hat{h}_{T+t'} &= \texttt{Back-AdaLN}_1(\mathcal{E}^{CA}_{T+t'}, \texttt{SelfAttn}) \\
    h'_{T+t'} &= \texttt{Back-AdaLN}_2(\hat{h}_{T+t'}, \texttt{GLU})
\end{aligned}
\end{equation}
Finally, a linear projector is used to recover the queried frames, $\mathcal{Y}_{T+t'} = \texttt{MLP}(h'_{T+t})$.

% Each future trigger ${\mathcal{E}}^{CA}_{T+t}$ is then fused with the normalized background to yield a context-aware representation:
% \begin{equation}
%     \mathcal{F}_{T+t'} = \texttt{Fuse}({\mathcal{E}}^{CA}_{T+t'}, \tilde{\mathcal{B}}_{T+t'}).
% \end{equation}
% To model cross-region interactions, we apply self-attention~\cite{vaswani2017attention} over $\mathcal{F}_{T+t}$:
% \begin{equation}
%     \hat{\mathcal{F}}_{T+t'} = \texttt{SelfAttn}(\mathcal{F}_{T+t'}). 
% \end{equation}
% Finally, a GLU refines spatially attended signals, followed by a linear projection to generate the predicted radar field:
% \begin{equation}
% \begin{aligned}
%     \mathcal{F}_{T+t'} &= \texttt{GLU}(\hat{\mathcal{F}}_{T+t'}), \\
%     \widehat{\mathcal{Y}}_{T+t'} &= \texttt{Linear}(\mathcal{F}_{T+t'}), \quad t' \in \{1, \dots, K\}.
% \end{aligned}
% \end{equation}

% 要改的现在只是用来占位置
\subsection{Training Objective}
% To promote physically meaningful and process-aware learning,  we have formulated a multi-channel loss that supervises each meteorological field using a separate MSE term. It is important to note that these fields are not equally informative; each one corresponds to a different stage of convective development:
% \begin{itemize}[leftmargin=*]
%     \item Kinematic variables ($AzShr$, $Div$, $SW$) reflect rapid low-level dynamics, such as convergence and shear, which often act as triggers for convection initiation.
%     \item Microphysical variables ($Z_H$, $K_{DP}$, $Z_{DR}$, $\rho_{HV}$) capture evolving storm structure, including hydrometeor concentration, phase changes, and particle alignment.
%     \item Thermodynamic features, embedded in the background field $\mathcal{B}$, modulate storm potential over mesoscale time horizons.
% \end{itemize}
% To reflect these differences, the loss function assigns channel-specific penalties, allowing the model to focus more on physically sensitive variables critical for storm nowcasting. Temporal consistency is enforced by aggregating error across all $T$ predicted steps:
% \begin{align}
% &\mathcal{L} = \sum_{c\in C} \cdot \sum_{t=T}^{T+K} | \hat{\mathbf{Y}}_t^c - \mathbf{Y}_t^c \|_2^2, \nonumber \\
% &\ C \in \{ {Z_{H},K_{DP}, Z_{DR}, r_{HV},Azhr, Div, SW} \}
% \end{align}

%损失函数纠正
To enable process-aware temporal learning, we formulate a multi-task loss comprising two components:
\begin{align}
\mathcal{L} = \sum_{t=T+1}^K(\mathcal{\hat{Y}}_{t} - \mathcal{X}_{t})^2 + \lambda_s\sum_{t=0}^T(\mathcal{\hat{Y}}_{t} - \mathcal{X}_{t})^2.
\end{align}
The first loss term supervises prediction of future frames, while the second (weighted by $\lambda_s$) reconstructs input frames to inject prior knowledge of convective evolution.
To reinforce this guidance early in training, we apply a warm-up phase with an increased $\lambda_s$ for the initial epochs.

%% file: sec/5experiments.tex
\section{Evaluation}
\label{sec:evaluation}

This section evaluates \N\ through five key questions (Implementation details in the Appendix~\ref{Implementation}.):
\begin{itemize}[leftmargin=*]
    \item \textbf{RQ1}: What is \N’s overall performance.
    \item \textbf{RQ2}: How robust is \N~ across lead times.
    \item \textbf{RQ3}: Performance during storm initiation and dissipation.
    \item \textbf{RQ4}: How does each component affect performance.
    \item \textbf{RQ5}: Physical realism and interpretability in case studies.
\end{itemize}
% RQ1:WhatisEchoPilot’soverallperformance? 314 •RQ2:HowrobustisEchoPilotacrossleadtimes? 315 •RQ3:HowdoesEchoPilotperformduringstormini-316 tiationanddissipationphases? 317 •RQ4:Howdoeseachcomponentaffectperformance? 318 •RQ5: DoesEchoPilotproducephysically realistic 319 andinterpretableforecastsinvisualizedcasestudies?

\subsection{Evaluation Settings}
% \subsubsection{Datasets}
\subsubsection{Datasets}
We evaluate \N\ on the \textbf{3D-NEXRDA} benchmark~\cite{Datasets}, which contains high-resolution 3D radar observations of severe convective storms across the continental United States (2020--2022). Each sample consists of 25 frames (125 minutes) at $2\,\text{km}$ spatial resolution with 28 vertical levels. The dataset includes 7 key radar variables, covering reflectivity ($Z_H$) and representative microphysical and kinematic fields. Additional details are provided in Appendix~\ref{sec:appendix_dataset} \& ~\ref{split}.

\subsubsection{Baselines}

We compare~\N\ against 11 competitive baselines, including
 \textbf{ConvLSTM}~\cite{shi2015convolutional}, \textbf{PhyDNet}~\cite{guen2020disentangling}, 
 \textbf{PredRNN-V2}~\cite{wang2022predrnn}, \textbf{Earthformer}~\cite{gao2022earthformer}, \textbf{Earthfarseer}~\cite{wu2024earthfarsser}, 
 \textbf{WF-UNet}~\cite{kaparakis2023wf}, \textbf{SimVP}~\cite{gao2022simvp}, \textbf{DGMR}~\cite{ravuri2021skilful}, \textbf{NowcastNet}~\cite{zhang2023skilful}, \textbf{PastNet}~\cite{wu2024pastnet},  and\textbf{AlphaPre}~\cite{Lin_2025_CVPR}
 (detailed in Appendix~\ref{baseline}).

\subsubsection{Metrics}
% To evaluate both the accuracy and physical realism of~\N's predictions, we employ a comprehensive set of metrics grouped into three categories:
% First, \textbf{Mean Absolute Error (MAE)} measures per-pixel numerical deviation, where \textit{lower values indicate more accurate forecasting}. 
% Second, \textbf{Critical Success Index (CSI)} is computed at 20, 30, and 40 dBZ thresholds to assess~\N's ability to detect storms of varying severity, where \textit{higher CSI values indicate improved detection of high-impact convective cores with fewer false alarms.}
% Third, \textbf{Structural Similarity Index (SSIM)} and \textbf{Signal-to-Noise Ratio (SNR)} evaluate the preservation of storm morphology and textural sharpness, where \textit{higher values imply more coherent and visually plausible storm structures.}

To evaluate both the accuracy and physical realism of~\N's predictions, we employ a comprehensive set of metrics: Mean Absolute Error (MAE), Critical Success Index (CSI) at various reflectivity thresholds, Structural Similarity Index (SSIM), and Signal-to-Noise Ratio (SNR). Detailed definitions and mathematical formulations for all metrics are provided in Appendix~\ref{appendix:metrics}.

\begin{table}[htbp]
\renewcommand{\arraystretch}{0.85}
\caption{Performance under Single-Radar Variable Outputs.}
\label{tab:SISO&MISO-performance}
\centering
\small
\setlength{\tabcolsep}{5pt}
\begin{tabular}{l ccccccc}
\toprule
\textbf{Model} & \textbf{MAE↓} & \textbf{mCSI↑} & \textbf{CSI$_{20}$↑} & \textbf{CSI$_{30}$↑} & \textbf{CSI$_{40}$↑} & \textbf{SNR↑} & \textbf{SSIM↑} \\
\midrule
\multicolumn{8}{c}{\textbf{Single-Input Single-Output (SISO)}} \\
\midrule
ConvLSTM       & 3.012 & 0.405 & 0.561 & 0.449 & 0.204 & 5.112 & 0.706 \\
PredRNN-V2     & 3.089 & 0.393 & 0.549 & 0.436 & 0.194 & 4.955 & 0.700 \\
PhyDNet        & 2.880 & 0.420 & 0.593 & 0.480 & 0.187 & 5.962 & 0.714 \\
Earthformer    & 3.002 & 0.402 & 0.597 & 0.473 & 0.137 & 6.259 & 0.675 \\
Earthfarseer   & 2.959 & 0.426 & 0.591 & 0.472 & 0.214 & 5.762 & 0.674 \\
SimVP          & 2.803 & 0.444 & 0.613 & 0.501 & 0.219 & 6.296 & 0.727 \\
PastNet        & 3.128 & 0.400 & 0.578 & 0.455 & 0.167 & 5.345 & 0.661 \\
AlphaPre       & 2.801 & 0.439 & 0.604 & 0.488 & 0.225 & 6.336 & 0.719 \\
DGMR           & 2.949 & 0.376 & 0.544 & 0.433 & 0.152 & 4.358 & 0.684 \\
NowcastNet     & 3.699 & 0.380 & 0.516 & 0.411 & 0.213 & 3.389 & 0.649 \\
\midrule
\multicolumn{8}{c}{\textbf{Multi-Input Single-Output (MISO)}} \\
\midrule
ConvLSTM       & 3.078 & 0.402 & 0.554 & 0.446 & 0.205 & 4.858 & 0.700 \\
PredRNN-V2     & 3.028 & 0.404 & 0.556 & 0.449 & 0.207 & 4.925 & 0.703 \\
PhyDNet        & 2.909 & 0.429 & 0.588 & 0.486 & 0.213 & 5.836 & 0.716 \\
Earthformer    & 2.909 & 0.442 & 0.614 & 0.505 & 0.208 & 6.389 & 0.688 \\
Earthfarseer   & 2.830 & 0.414 & 0.600 & 0.461 & 0.181 & 6.152 & 0.693 \\
SimVP          & 3.000 & 0.429 & 0.576 & 0.475 & 0.237 & 5.282 & 0.706 \\
PastNet        & 3.279 & 0.368 & 0.571 & 0.426 & 0.108 & 5.461 & 0.596 \\
AlphaPre       & 2.911 & 0.441 & 0.606 & 0.491 & 0.227 & 6.209 & 0.705 \\
WF-UNet        & 2.897 & 0.398 & 0.578 & 0.448 & 0.169 & 5.929 & 0.712 \\
DGMR           & 2.780 & 0.340 & 0.556 & 0.402 & 0.062 & 5.033 & 0.710 \\
NowcastNet     & 3.899 & 0.326 & 0.468 & 0.350 & 0.162 & 3.143 & 0.607 \\
\midrule
\textbf{Ours}  & \textbf{2.622} & \textbf{0.464} & \textbf{0.616} & \textbf{0.514} & \textbf{0.260} & \textbf{6.544} & \textbf{0.746}\\
\midrule
\textit{Improve.(\%)} & \textbf{7.349} & \textbf{4.973} & \textbf{0.326} & \textbf{1.782} & \textbf{9.705} & \textbf{2.426} & \textbf{4.195} \\
\bottomrule
\end{tabular}
\end{table}

\begin{table}[htbp]
\renewcommand{\arraystretch}{0.85}
\caption{Performance under Multi-Radar Variable Outputs (\textbf{MIMO}).}
\label{tab:MIMO-performance}
\centering
\small
\setlength{\tabcolsep}{5pt}
\begin{tabular}{l ccccccc}
\toprule
\textbf{Model} & \textbf{MAE↓} & \textbf{mCSI↑} & \textbf{CSI$_{20}$↑} & \textbf{CSI$_{30}$↑} & \textbf{CSI$_{40}$↑} & \textbf{SNR↑} & \textbf{SSIM↑}  \\
\midrule
ConvLSTM     & 3.197 & 0.368 & 0.527 & 0.413 & 0.163 & 4.348 & 0.681 \\
PredRNN-V2   & 2.975 & 0.354 & 0.547 & 0.403 & 0.112 & 5.198 & 0.697 \\
PhyDNet      & 3.967 & 0.313 & 0.528 & 0.367 & 0.045 & 4.113 & 0.566 \\
Earthformer  & 3.192 & 0.324 & 0.552 & 0.392 & 0.027 & 4.947 & 0.679 \\
Earthfarseer & 3.023 & 0.358 & 0.586 & 0.414 & 0.074 & 5.514 & 0.669 \\
SimVP        & 2.717 & 0.407 & \textbf{0.607} & 0.472 & 0.144 & 6.076 & 0.727  \\
PastNet      & 3.019 & 0.350 & 0.566 & 0.405 & 0.079 & 5.208 & 0.676 \\
AlphaPre     & 3.177 & 0.392 & 0.577 & 0.435 & 0.162 & 5.742 & 0.658  \\
WF-UNet      & 2.981 & 0.376 & 0.571 & 0.436 & 0.120 & 5.393 & 0.701 \\
DGMR         & 2.947 & 0.272 & 0.492 & 0.295 & 0.029 & 4.257 & 0.691 \\
NowcastNet   & 4.497 & 0.371 & 0.498 & 0.404 & 0.212 & 2.092 & 0.639 \\
\midrule
\textbf{Ours} & \textbf{2.660} & \textbf{0.431} & \textbf{0.607} & \textbf{0.488} & \textbf{0.198} & \textbf{6.183} & \textbf{0.738} \\
\midrule
\textit{Improve.(\%)} & \textbf{2.096} & \textbf{5.900} & \textbf{0.000} & \textbf{3.390} & \textbf{21.472} & \textbf{1.761} & \textbf{1.513} \\
\bottomrule
\end{tabular}
\vspace{-1em}
\end{table}

\subsection{Overall Performance (\textbf{RQ1})}

We evaluate~\N\ under two setups to reflect operational needs and physical complexity:
\begin{itemize}[leftmargin=*]
    \item \textbf{Single-Output (SO):} Predicting only reflectivity ($Z_H$) using either single (SISO) or multi-source (MISO) inputs. This assesses whether auxiliary physical fields improve $Z_H$ forecasting (Table~\ref{tab:SISO&MISO-performance}).
    \item \textbf{Multi-Output (MO):} Jointly forecasting $Z_H$ alongside six microphysical ($K_{DP}$, $Z_{DR}$, $r_{HV}$) and kinematic ($Az_{hr}$, $Div$, $SW$) fields. This evaluates~\N's physical consistency and generalization under full-field forecasting (Tables~\ref{tab:MIMO-performance} and \ref{tab:combined_horizontal}).
\end{itemize}
For $Z_H$, we report MAE, structure-aware (SSIM, SNR), and threshold-based (CSI$_{20/30/40}$, mCSI) metrics. For auxiliary fields, we focus on SNR and SSIM to measure signal fidelity and spatial coherence.

\textbf{SO Forecasting.}
Table~\ref{tab:SISO&MISO-performance} reveals three key findings under the SISO and MISO settings:
\begin{itemize}[leftmargin=*]
    % \item \textbf{SISO Architecture Constraints:} Recurrent models (e.g., \textit{ConvLSTM}) suffer from over-smoothing, yielding lower mCSI and SNR. \textit{PhyDNet} mitigates this via a physics-guided inductive bias. Conversely, recurrent-free models (e.g., \textit{AlphaPre}, \textit{SimVP}) better preserve spatial details, though Fourier-based models (\textit{Earthfarseer}, \textit{PastNet}) struggle with high-intensity peaks (CSI$_{40}$).
    \item \textbf{MISO: Response varies by architecture.}
    Most recurrent models degrade or stagnate under MISO (e.g., \textit{ConvLSTM}), reflecting limited capacity for fusing heterogeneous physical inputs.
    In contrast, \textit{AlphaPre} and \textit{Earthformer} improve consistently, suggesting that transformer and CNN backbones offer better spatial alignment and cross-channel disentanglement. 
    However, \textit{fourier-based models} (\textit{Earthfarseer, PastNet}) degrade in CSI$_{40}$, likely due to low-frequency bias that limits peak recovery. 
    \item \textbf{MISO: Response varies by architecture:} While most recurrent models stagnate or degrade with multi-source inputs, CNN and transformer backbones (\textit{AlphaPre}, \textit{Earthformer}) consistently improve, indicating superior cross-channel disentanglement.
    \item \textbf{\N's Superiority:} \N\ outperforms all baselines in both settings. Compared to the second-best models, it improves CSI$_{40}$ by up to 16.03\%, mCSI by 6.10\%, and reduces MAE by 4.05\%, demonstrating highly effective multimodal fusion for severe weather forecasting.
\end{itemize}

\textbf{MO Forecasting.}
Jointly modeling heterogeneous physical fields challenges most baselines, leading to degraded structure-sensitive metrics (e.g., CSI$_{40}$ drops by 44\%--81\% for \textit{PastNet} and \textit{PhyDNet}). This reflects the difficulty of balancing conflicting dynamics, such as smooth microphysical fields versus sharp kinematic gradients. 
In contrast, \N\ maintains robust performance across all metrics, achieving a 21.47\% gain in CSI$_{40}$, a 1.76 dB increase in SNR, and the highest mCSI for $Z_H$ (Table~\ref{tab:MIMO-performance}). Furthermore, as shown in Table~\ref{tab:combined_horizontal}, \N\ achieves best or near-best SNR and SSIM on the six auxiliary variables, proving its capability for physically consistent, full-spectrum storm diagnosis.

\vspace{-1em}
\begin{table}[h]
\renewcommand{\arraystretch}{0.85}
\caption{Performance comparison across Microphysical and Kinematic Variables.}
\label{tab:combined_horizontal}
\centering
\resizebox{\linewidth}{!}{
\setlength{\tabcolsep}{2pt}
\begin{tabular}{l cccccc cccccc}
\toprule
\multirow{3}{*}{\textbf{Method}} & \multicolumn{6}{c}{\textbf{Microphysical Variables}} & \multicolumn{6}{c}{\textbf{Kinematic Variables}} \\
\cmidrule(lr){2-7} \cmidrule(lr){8-13}
 & \multicolumn{2}{c}{$K_{DP}$} & \multicolumn{2}{c}{$Z_{DR}$} & \multicolumn{2}{c}{$r_{HV}$} & \multicolumn{2}{c}{$AzShr$} & \multicolumn{2}{c}{$Div$} & \multicolumn{2}{c}{$SW$} \\
\cmidrule(lr){2-3} \cmidrule(lr){4-5} \cmidrule(lr){6-7} \cmidrule(lr){8-9} \cmidrule(lr){10-11} \cmidrule(lr){12-13}
 & \textbf{SNR↑} & \textbf{SSIM↑} & \textbf{SNR↑} & \textbf{SSIM↑} & \textbf{SNR↑} & \textbf{SSIM↑} & \textbf{SNR↑} & \textbf{SSIM↑} & \textbf{SNR↑} & \textbf{SSIM↑} & \textbf{SNR↑} & \textbf{SSIM↑} \\
\midrule
ConvLSTM     & 34.031 & 0.976 & 24.063 & 0.882 & 12.243 & 0.594 & 28.917 & 0.932 & 25.875 & 0.893 & 3.059 & 0.679 \\
PredRNN-V2   & 33.354 & 0.971 & 24.453 & 0.889 & 12.991 & 0.618 & 28.902 & 0.932 & 25.916 & 0.894 & 3.606 & 0.697 \\
PhyDNet      & 33.590 & 0.976 & 24.424 & 0.896 & 13.136 & 0.629 & 28.841 & 0.932 & 25.640 & 0.891 & 2.892 & 0.600 \\
Earthformer  & 33.732 & 0.975 & 24.647 & 0.896 & 13.262 & 0.631 & 28.743 & 0.932 & 25.612 & 0.891 & 3.516 & 0.687 \\
Earthfarseer & 32.171 & 0.974 & 24.519 & 0.897 & 13.758 & 0.649 & 28.112 & 0.930 & 25.220 & 0.890 & 3.633 & 0.689 \\
SimVP        & 33.961 & 0.976 & 24.830 & 0.899 & 13.721 & 0.660 & 28.956 & \textbf{0.934} & 26.111 & 0.898 & 3.817 & 0.711 \\
PastNet      & 33.196 & 0.975 & 24.504 & 0.893 & 13.400 & 0.626 & 28.551 & 0.931 & 25.445 & 0.890 & 3.676 & 0.670 \\
AlphaPre     & 33.694 & 0.975 & 34.839 & 0.900 & 13.727 & 0.651 & 28.735 & 0.931 & 25.847 & 0.894 & 4.067 & 0.676 \\
WF-UNet      & 34.030 & 0.976 & 24.814 & \textbf{0.900} & 13.578 & 0.647 & 28.878 & 0.932 & \textbf{26.148} & \textbf{0.898} & 4.114 & 0.715 \\
DGMR         & 33.604 & 0.973 & 23.835 & 0.890 & 12.332 & 0.627 & 28.724 & 0.930 & 25.572 & 0.889 & 3.042 & 0.691 \\
NowcastNet   & 30.260 & 0.961 & 21.223 & 0.862 & 9.402 & 0.515 & 24.973 & 0.909 & 21.961 & 0.863 & 0.569 & 0.626 \\
\midrule
\textbf{Ours} & \textbf{34.057} & \textbf{0.976} & \textbf{24.936} & 0.899 & \textbf{13.798} & \textbf{0.662} & \textbf{28.984} & 0.933 & 26.099 & 0.896 & \textbf{4.240} & \textbf{0.725} \\
\bottomrule
\end{tabular}
}
\vspace{-1em}

\end{table}

% \begin{table}[h]
% \centering
% \resizebox{\linewidth}{!}{
% \begin{tabular}{lccccccc}
% \toprule
% \textbf{CAVR Phase} & \textbf{Models} & \textbf{MAE$\downarrow$} & \textbf{mCSI$\uparrow$} & \textbf{CSI_{20}$\uparrow$} & \textbf{CSI_{30}$\uparrow$} & \textbf{CSI_{40}$\uparrow$} & \textbf{SSIM$\uparrow$} \\
% \midrule
% \multirow{3}{*}{\multicolumn{1}{c}{CAVR $>$ 0}}
%  & SimVP & 3.399 & 0.420 & 0.600 & 0.464 & 0.146 & 0.669 \\
%  & WF-UNet & 3.439 & 0.405 & 0.584 & 0.463 & 0.169 & 0.675 \\
%  & \textbf{Ours}  & \textbf{3.188} & \textbf{0.441} & \textbf{0.616} & \textbf{0.508} & \textbf{0.201} & \textbf{0.706} \\
% \midrule
% \multirow{3}{*}{\multicolumn{1}{c}{CAVR $<$ 0}}
%  & SimVP & 2.818 & 0.421 & 0.603 & 0.509 & 0.152 & 0.728 \\
%  & WF-UNet &  2.994 & 0.411 & 0.577 & 0.487 & 0.169 & 0.706 \\
%  & \textbf{Ours}  & \textbf{2.711} & \textbf{0.442} & \textbf{0.620} & \textbf{0.514} & \textbf{0.210} & \textbf{0.733} \\
% \bottomrule
% \end{tabular}
% }
% \caption{Model performance under different convective phases.}
% \label{tab:cavr_phase}
% \end{table}

\subsection{Forecast Horizon Analysis (\textbf{RQ2})}
\begin{figure}[h]
\centering
% \caption{Sensitivity Analysis}
\includegraphics[width=0.7\linewidth]{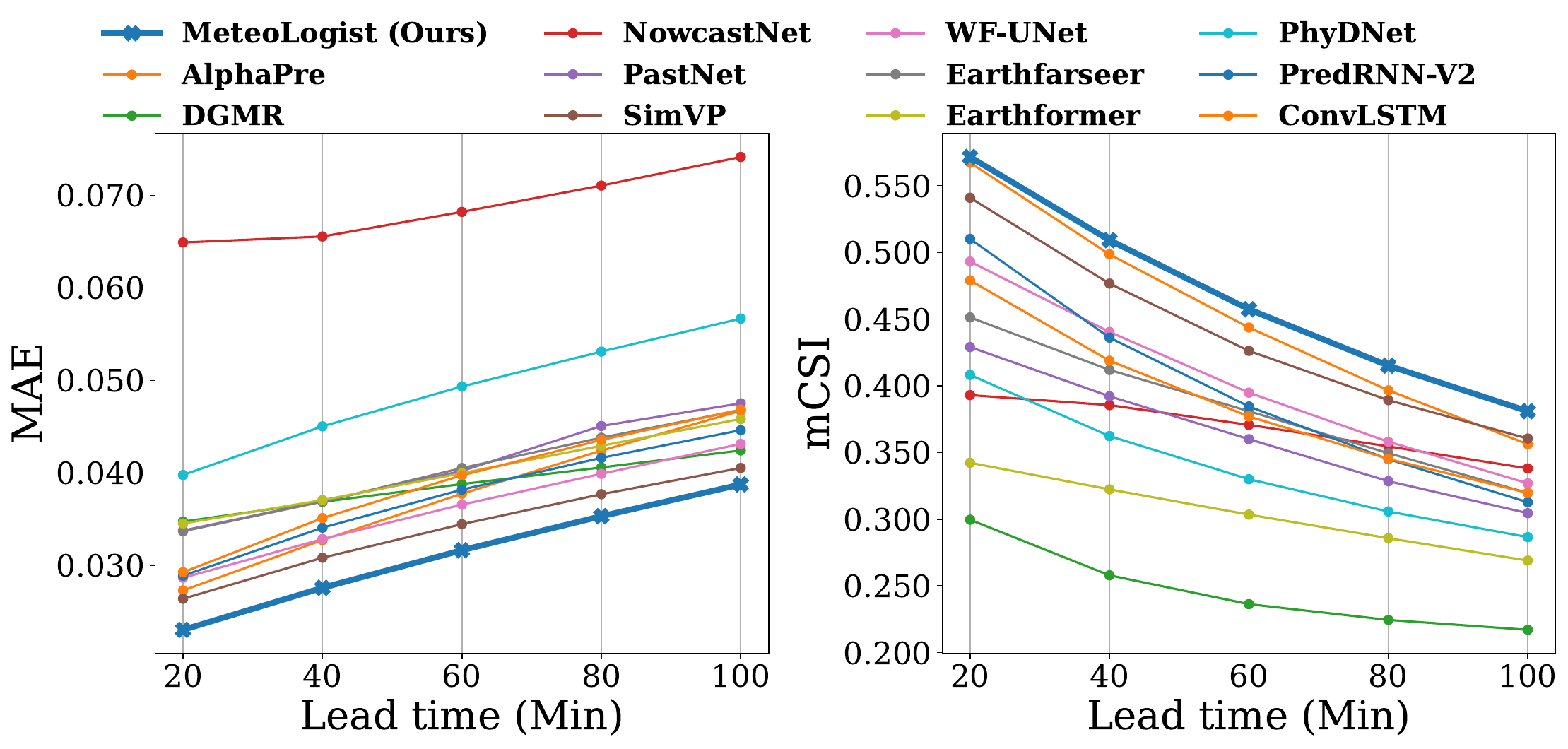}
\caption{
mCSI/MAE Trends across Lead Times  
}
\label{fig:sensitivity}
\vspace{-1em}
\end{figure}

% Figure~\ref{fig:sensitivity} illustrates model performance across forecast horizons from 20 to 100 minutes, using both pix-level error (MAE) and structure-aware accuracy (mCSI). 
Figure~\ref{fig:sensitivity} evaluates pixel-level error (MAE) and structural accuracy (mCSI) from 20 to 100 minutes forecast horizons.
As expected, performance degrades as lead time increases, but the rate of decline varies:
recurrent-based models (e.g., ConvLSTM, PredRNN-V2) deteriorate faster (mCSI: –0.022) than recurrent-free ones (–0.014), reflecting weaker spatial generalization over longer horizons.  
\N\ shows a slower decay, which demonstrates its forward-looking capability:
sensing convective organization well before conventional echo-based models fail.

% Figure~\ref{fig:sensitivity} evaluates pixel-level error (MAE) and structural accuracy (mCSI) from 20 to 100 minutes forecast horizons. While performance naturally degrades over time, recurrent models (e.g., ConvLSTM, PredRNN-V2) deteriorate faster (mCSI: –0.022) than recurrent-free baselines (–0.014), reflecting weaker spatial generalization. In contrast, \N\ exhibits the slowest decay, demonstrating its superior ability to anticipate convective organization well before conventional echo-based models fail.

\subsection{Development and Dissipation Analysis (\textbf{RQ3})}
To examine~\N's ability to model storm evolution, we evaluate it across the developing and dissipating life-cycle stages, identified via the Convective Area Variation Rate (CAVR)~\cite{pan2021improving} for echo growth (CAVR $>$ 0) and decay (CAVR $<$ 0). 
As summarized in Table~\ref{tab:cavr_phase},~\N\ consistently improves performance across both phases:
\begin{itemize}[leftmargin=*]
    \item \textbf{Developing Stage (CAVR $>$ 0):} \N\ achieves consistent gains across all metrics. Notably, CSI$_{40}$ increases by 37.67\% (from 0.146 to 0.201), indicating sharper detection of emerging convective cores that baselines tend to blur.
    \item \textbf{Dissipating Stage (CAVR $<$ 0):} \N\ effectively preserves structural details and limits spatial drift, yielding higher SSIM (0.733 vs. 0.728) and lower MAE (2.711 vs. 2.818).
\end{itemize}
These results demonstrate that integrating physical cues enables~\N\ to accurately capture both the intensification and decay processes of convective systems.

\vspace{-1em}
\begin{table}[h]
\centering
\caption{Model Performance under Different Convective Phases.}
\resizebox{0.8\linewidth}{!}{%
\begin{tabular}{lccccccc}
\toprule
\textbf{CAVR Phase} & \textbf{Models} & \textbf{MAE$\downarrow$} & \textbf{mCSI$\uparrow$} & \textbf{CSI$_{20}$ $\uparrow$} & \textbf{CSI$_{30}$ $\uparrow$} & \textbf{CSI$_{40}$ $\uparrow$} & \textbf{SSIM$\uparrow$} \\
\midrule
\multirow{4}{*}{\shortstack{CAVR $>$ 0}}
  & SimVP & 3.399 & 0.420 & 0.600 & 0.464 & 0.146 & 0.669 \\
 & WF-UNet & 3.482 & 0.412 & 0.588 & 0.470 & 0.178 & 0.671 \\
 & AlphaPre & 3.586 & 0.423 & \textbf{0.624}  & 0.487 & 0.156 & 0.603 \\
 & \textbf{Ours} & \textbf{3.249} & \textbf{0.446} & 0.620& \textbf{0.512} & \textbf{0.206} & \textbf{0.702} \\
  % & \textbf{Ours}  & \textbf{3.188} & \textbf{0.441} & \textbf{0.616} & \textbf{0.508} & \textbf{0.201} & \textbf{0.706} \\
\midrule
\multirow{4}{*}{\shortstack{CAVR $<$ 0}}
& SimVP & 2.890 & 0.432 & 0.604 & 0.530 & 0.162 & 0.725 \\
 & WF-UNet & 3.117 & 0.418 & 0.577 & 0.500 & 0.175 & 0.697 \\
 & AlphaPre & 3.203 & 0.453 & \textbf{0.616} & 0.527 & 0.196 & 0.629 \\
 & \textbf{Ours} & \textbf{2.835} & \textbf{0.454} & 0.614 & \textbf{0.528} & \textbf{0.220} & \textbf{0.728} \\
\bottomrule
\end{tabular}%
}
\vspace{-1em}
% \vspace{-1em}
\label{tab:cavr_phase}
\end{table}

\subsection{Key Components Assessment (\textbf{RQ4})}

Table~\ref{tab:ablation} quantifies the contribution of the microphysical and kinematic encoders (Further ablations regarding background information and temporal lead-lag stages are detailed in Appendix~\ref{sec:additional_ablation}.). 

\begin{itemize}[leftmargin=*]
    \item \textbf{Warmup Strategy (w/o Warmup):} Omitting the warmup degrades performance, particularly on high-intensity events (CSI$_{40}$ $\downarrow$ 12.12\%), emphasizing the necessity of progressive supervision.
    
    \item \textbf{Microphysical Encoder (w/o Micro):} Removing this yields significant drops in CSI$_{40}$ ($\downarrow$ 14.14\%), confirming that variables like $K_{DP}$ and $Z_{DR}$ are crucial for modeling intense precipitation.
    
    \item \textbf{Kinematic Encoder (w/o Kin):} Omission reduces CSI$_{40}$ ($\downarrow$ 9.10\%) and SSIM ($\downarrow$ 0.68\%), indicating that wind-convergence features (e.g., azimuthal shear) help sharpen convective edges.
    
    \item \textbf{Synergistic Effect (w/o Factors):} Removing both encoders causes the sharpest overall decline (e.g., CSI$_{40}$ $\downarrow$ 18.69\%), proving that microphysical and kinematic streams provide complementary insights critical for realistic storm reconstruction.
\end{itemize}

\vspace{-1.5em}
\begin{table}[h]
\centering
\caption{Ablation Study of Key Components.}
\label{tab:ablation}
\small
\setlength{\tabcolsep}{4pt}
\renewcommand{\arraystretch}{0.95}
\begin{tabular}{lccccccc}
\toprule
\textbf{Variant} & \textbf{MAE$\downarrow$} & \textbf{mCSI$\uparrow$} & \textbf{CSI$_{20}$$\uparrow$} & \textbf{CSI$_{30}$$\uparrow$} & \textbf{CSI$_{40}$$\uparrow$} & \textbf{SNR$\uparrow$} & \textbf{SSIM$\uparrow$} \\
\midrule
w/o Warmup  & 2.817 & 0.406 & 0.585 & 0.459 & 0.174 & 5.931 & 0.728 \\
w/o Micro   & 2.761 & 0.419 & 0.605 & 0.481 & 0.170 & 6.045 & 0.728 \\
w/o Kin     & 2.714 & 0.425 & 0.607 & 0.487 & 0.180 & 6.166 & 0.733 \\
w/o Factors & 2.746 & 0.409 & 0.602 & 0.464 & 0.161 & 6.123 & 0.729 \\
\midrule
\textbf{Ours} & \textbf{2.660} & \textbf{0.431} & \textbf{0.607} & \textbf{0.488} & \textbf{0.198} & \textbf{6.183} & \textbf{0.738} \\
\bottomrule
\end{tabular}
\vspace{-1em}
\end{table}

% illustrates the temporal evolution of two key performance metrics—mCSI and perceptual similarity measured by MAE. 
% These curves provide insight into how different models perform as the forecast horizon increases. 
% As expected, performance deteriorates across all methods with longer lead times due to increased uncertainty. 
% However, our method consistently maintains a lower MAE compared to others, suggesting superior robustness to temporal degradation—a common challenge in long-range forecasting.
% In terms of mCSI, which evaluates both the spatial alignment and intensity accuracy of rainfall predictions, our model again demonstrates strong performance. 
% Notably, it achieves consistently higher mCSI scores across time steps than most baselines, indicating its effectiveness in preserving the meteorological structure and intensity distribution of precipitation events. 
% This highlights our model's capability to mitigate the blurring and underestimation issues typically observed in long-sequence predictions.

% \begin{figure}[h]
% \centering
% % \caption{Sensitivity Analysis}
% \includegraphics[width=1\linewidth]{figures/CAVR.pdf}
% \caption{
% Model performance under different convective phases. }
% \label{fig:cavr_phase}
% % \vspace{-0.2in}
% \end{figure} 

\subsection{Case Study (\textbf{RQ5})}

Figure~\ref{fig:case} in Appendix~\ref{vis} shows a visual comparison of models for a 100-minute radar forecasting. 
Baselines tend to blur fine-scale convection or misplace rainfall centers as lead time increases, obscuring how storms grow.
In contrast,~\N\ clearly delineates the growth trajectory of convective systems—capturing the gradual intensification and spatial expansion of rain cells with remarkable fidelity.
Even at 100 minutes, it reconstructs both the morphology and intensity of evolving convective cores, closely matching ground truth (GT).
These results highlight \N’s ability to perceive and preserve storm growth dynamics across extended horizons—transforming forecasting from passive extrapolation to physically coherent evolution modeling.

%% file: sec/7conclusion.tex
\section{Conclusion}
\label{sec:conclusion}

In this study, we present~\N, a physics-inspired framework that reframes echo extrapolation from tracking echoes to foreseeing convective birth.
\N\ tackles two core challenges in early-stage convection modeling: 
 (1) the temporal desynchronization of thermodynamic, kinematic, and microphysical drivers, 
 and (2) the spatial disorganization of convective energy.
To this end,~\N\ introduces three physically grounded components:
i)~\ComponentA\ disentangle radar inputs by meteorological processes,
ii)~\ComponentB\ infers trigger timing across modalities,
and ii)~\ComponentC\ integrates non-local cues to restore spatial coherence. 
Evaluated on the large-scale 3D-NEXRAD benchmark (2020–2022, US-wide), \N\ sets a new state of the art, boosting CSI$_{40}$ by up to 16.03\%, mCSI by 6.10\%, and reducing MAE by 4.05\%. 
\textbf{Beyond accuracy,~\N\ marks a conceptual shift—from extrapolating what we see, to reasoning about what the atmosphere is about to do.}